\documentclass[twocolumn,superscriptaddress,prl] {revtex4}
\usepackage{graphicx}
\usepackage{amsmath}
\usepackage{amssymb}
\usepackage{amscd}
\usepackage{color}
\usepackage{pstricks}
\usepackage{psfrag}
\usepackage[pdftitle={Event-chain algorithms for hard-sphere systems},
            pdfauthor={Etienne P. Bernard, Werner Krauth, and David B. Wilson},
            breaklinks=true,
            bookmarks=true,bookmarksopen=true,bookmarksopenlevel=2]{hyperref}
\usepackage[all]{hypcap}
\usepackage{breakurl}

\newcommand{\Dloc}{D_\text{loc}}
\newcommand{\Delt}{\Delta_t}
\newcommand{\Delx}{\Delta_\xvec}
\newcommand{\SEC}{SEC}
\newcommand{\REC}{REC}
\newcommand{\acc}{\text{acc}}
\newcommand{\flabel}[1]{\label{f:#1}}
\newcommand{\elabel}[1]{\label{e:#1}}

\newcommand{\tlabel}[1]{\label{t:#1}}
\newcommand{\eq}[1]{Eq.~(\ref{e:#1})}

\newcommand{\fig}[1]{\hyperref[f:#1]{Fig.~\ref*{f:#1}}}
\newcommand{\figg}[1]{\hyperref[f:#1]{Figure~\ref*{f:#1}}}
\newcommand{\tab}[1]{\hyperref[t:#1]{Table~\ref*{t:#1}}}

\newcommand{\quot}[1]{``#1''}

\newcommand{\VEC}[1]{\mathbf{#1}}
\newcommand{\xvec}{\VEC{x}}
    
\newcommand{\expb}[1]{\exp \glb #1 \grb} 

\newcommand{\glb}{\left(}  
\newcommand{\grb}{\right)}  

\newcommand{\dd}[1]{\text{d}{#1\ }}   

\newcommand{\mean}[1]{\left\langle #1 \right\rangle}
\begin{document}
\title{Event-chain Monte Carlo algorithms for hard-sphere systems
\footnote{This research was initiated at the \href{http://www.kitp.ucsb.edu/}{Kavli
Institute for Theoretical} \href{http://www.kitp.ucsb.edu/}{Physics} at UCSB. We thank the NSF for partial support under grant number PHY05-51164. We are grateful to M. Isobe for helpful correspondence.}}

\author{Etienne P. Bernard}
\email{etienne.bernard@ens.fr}
\author{Werner Krauth}
\email{werner.krauth@ens.fr}
\affiliation{CNRS-Laboratoire de Physique Statistique, Ecole Normale
Sup\'{e}rieure, 24 rue Lhomond, 75231 Paris Cedex 05, France}
\author{David B. Wilson} 
\homepage{http://dbwilson.com}
\affiliation{Microsoft Research, One Microsoft Way,
Redmond, Washington 98052, USA}
\date{May 01, 2009}
\begin{abstract}
In this paper we present the event-chain algorithms, which are fast
Markov-chain Monte Carlo methods for hard spheres and related systems.
In a single move of these rejection-free methods, an arbitrarily long
chain of particles is displaced, and long-range coherent motion can
be induced.  Numerical simulations show that event-chain algorithms
clearly outperform the conventional Metropolis method. Irreversible
versions of the algorithms, which violate detailed balance, improve the
speed of the method even further. We also compare our method with
a recent implementations of the molecular-dynamics algorithm. 
\end{abstract}

\maketitle

Hard spheres in three and in two dimensions (hard disks) occupy a special
place in statistical mechanics. Indeed, many fundamental concepts, from
the virial expansion (by van der Waals and Boltzmann), to two-dimensional
melting \cite{alder_phase}, to long-time tails \cite{alder_long_time},
were first discussed in these extraordinarily rich physical systems.
These models have also played a crucial role in the history of
computation: both the Metropolis algorithm \cite{metropolis} and molecular
dynamics \cite{alder} were first implemented for monodisperse hard disks
in a box. In contrast with the spectacular algorithmic developments
for lattice spin models \cite{swendsen,wolff}, Monte Carlo algorithms for
hard spheres have changed little since the 1950s, especially for high
densities. Recent sophisticated implementation have reduced the
complexity of the molecular dynamics algorithm to a value comparable to
that of the Monte Carlo method. Nevertheless, one can today
still not equilibrate sufficiently large systems \cite{mak} to clarify
whether the melting transition in two-dimensional hard disks, at density
(occupied volume fraction) $\eta \simeq 0.70$, is weakly first order,
or whether it is of the Kosterlitz-Thouless type \cite{nelson}, with a
narrow hexatic phase in between the liquid and the solid.

In this paper, we propose a class of Monte Carlo algorithms for
hard-sphere systems: the \quot{event-chain} algorithms. In contrast
to the Metropolis algorithm, these methods are rejection-free. In a single move, they
displace an arbitrary long chain of spheres, each advancing until it
strikes the next one.  Event-chain algorithms are generically faster
than other Markov-chain algorithms, in part because the mean-square
displacements of individual particles are larger.  In addition, one of the
event-chain algorithms moves coherently over long distances. This further
improves equilibration times.  Finally, the absence of rejections allows
us to consider irreversible versions, which violate detailed balance, but
preserve the correct stationary distribution. These versions accelerate
the algorithm even further.The event-chain algorithms clearly
outperform the traditional Metropolis algorithm for hard-disk and
hard-sphere systems.

\begin{figure}[b]
  \vspace*{-6pt}
  \centerline{\includegraphics{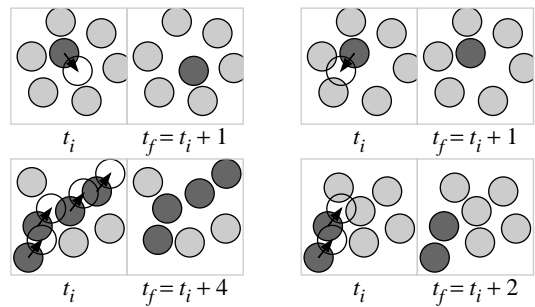}}
  \caption{
    \emph{Upper panels}: Accepted (\emph{left}) and rejected
    (\emph{right}) local Metropolis moves of a disk in the cage formed
    by its neighbors.
    \emph{Lower panels}: Accepted and rejected moves in Jaster's chain 
    algorithm.}
  \label{f:markov_moves_jaster_moves}
\end{figure}

In the local Metropolis algorithm, the move of a sphere is
accepted if it induces no overlaps, and is rejected otherwise (see
\fig{markov_moves_jaster_moves}).  This algorithm is notoriously slow
at high density because, although a particle can move back and forth in
the \quot{cage} formed by its neighbors, it cannot easily escape from
it \cite{SMAC}.

To overcome the limitations of the local Metropolis algorithm, coordinated
particle moves have been considered: When the displacement of one sphere
generates overlaps with other spheres, the latter are in turn moved
out of the way. The rejection-free pivot cluster algorithm
\cite{dress}, for example, works extremely well for binary \cite{buhot}
or for polydisperse \cite{santen} mixtures, but it breaks down for the
high densities of interest in two-dimensional melting.
In Jaster's algorithm \cite{jaster}, overlapping spheres forming a chain
are displaced, all of them by a fixed vector, until a configuration without
overlaps is obtained (see \fig{markov_moves_jaster_moves}). If a sphere branches out to more than one other
sphere during the chain construction, the move is rejected (see \fig{markov_moves_jaster_moves}).
This happens frequently, so the expected chain length 
is short and Jaster's algorithm barely faster than
the local Metropolis algorithm. 
\begin{figure}[htbp]
   \centerline{\includegraphics{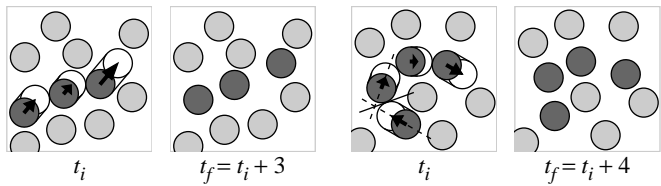}}
   \caption
      {\emph{Left two panels}: Move of the straight event-chain (\SEC)
      algorithm.  The individual  displacements add up to a distance
      $\ell$. \emph{Right two panels}: Reflected event-chain (\REC) move.}
   \label{f:event_chain_moves}
\end{figure}

In the algorithms presented here, each move consists in a deterministic
chain of \quot{events}: a disk advances until it strikes another one,
which then in turn is displaced. The move starts with a randomly chosen
disk, and stops when the lengths of all displacements add up to a
total-displacement parameter $\ell$ (see \fig{event_chain_moves}). This
parameter allows the move to be reversible without rejections. To satisfy
detailed balance, the move must also conserve configuration-space volume.
This implies that when a disk strikes a neighbor, the latter may be
displaced either in the original direction (\quot{straight event-chain}
(\SEC) algorithm) or in the direction reflected with respect to the
symmetry axis of the collision (\quot{reflected event-chain} (\REC)
algorithm) (see \fig{event_chain_moves}). In a periodic box, and with
the initial direction $\theta$ sampled uniformly in $[0,2\pi]$, both
versions, which we call \quot{reversible}, preserve the uniform measure
because of detailed balance.

\begin{figure}[htbp]
   \centerline{\includegraphics{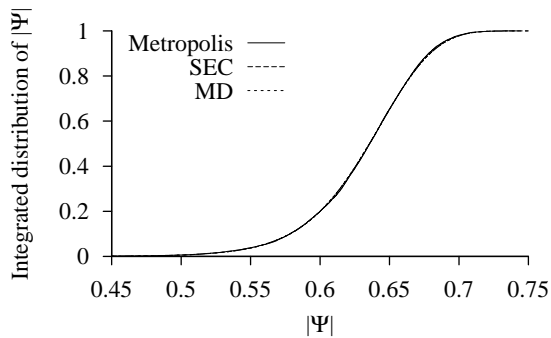}}
   \caption {Comparison of the integrated distribution of an
   observable (the absolute value of the order parameter  $\Psi$
   of \eq{order_parameter}) between the \SEC\ algorithm which breaks
   detailed balance, molecular dynamics (MD) and the local Metropolis
   algorithm for 1024 disks at $\eta=0.71$.}
   \flabel{distrib}
\end{figure}

The detailed balance condition is allowed to be broken in the
\SEC\ algorithm. Indeed, for a given direction
$\theta$, a configuration $\Gamma$ of $N$ disks can reach $N$ other
configurations in one move. By applying to $\Gamma$ the $N$ possible
moves in direction $-\theta$, one finds the $N$ configurations that reach
$\Gamma$. Therefore,  the \SEC\ algorithm satisfies global balance for
any distribution of $\theta$.  Algorithms breaking detailed balance
induce probability flows in the configuration space and potentially
speed-up equilibration\cite{MR1789978}. We study such an irreversible
version of the \SEC\ algorithm where $\theta$ is uniformly distributed in
$[0,\pi]$. To preserve ergodicity, at least two independent directions
are needed. By far our fastest implementation (the \quot{$xy$ version}
of the \SEC\ algorithm) alternates moves in the positive $x$ and $y$
directions ($\theta=0, \pi/2$).  A version of the \SEC\ algorithm,
but with rejections and which cannot break detailed balance, was also
mentioned in \cite{jaster}.

In \fig{distrib} we show the integrated distribution of $| \Psi |$ of
\eq{order_parameter}
\begin{equation}
\int_0^x  \pi(| \Psi |)  \dd  | \Psi |  
\end{equation}
for the $xy$ version of the \SEC\ algorithm, for the Metropolis algorithm
and for molecular dynamics in the same system. They are found to
be equal. This demonstrates the correctness of our implementations.

As a first step to analyze the performance of the event-chain algorithms,
we consider the mean-square displacement $\mean{\Delx^2}$ of individual
disks, both in the local Metropolis and in the event-chain algorithms.
As mentioned, event-chain algorithms generically outperform the Metropolis
algorithm in part because they take larger steps on average.  In order
to compare the two methods, we measure time in units of attempted
one-particle displacements.

\begin{figure}[htbp]
   \centerline{\includegraphics{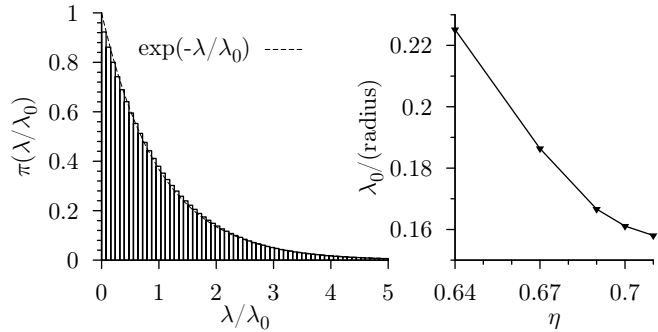}}
   \caption
   {\emph{Left}: Histogram $\pi(\lambda/\lambda_0)$
of the free path $\lambda(\theta)$ for $1024$ disks at
density $\eta=0.71$. The distribution is close to  exponential even in
the solid phase. \emph{Right}: Mean-free path $\lambda_0$
in units of the disk radius as a function of $\eta$.}
   \label{f:freepath}
\end{figure}

Let us define the \quot{free path} $\lambda = \lambda(\theta)$ of a disk
as the distance it must move in direction $\theta$ to strike another
particle. The ensemble average of $\lambda$ yields the mean-free path
$\lambda_0$.  The distribution of the free path $\pi(\lambda/\lambda_0)$
is well approximated by an exponential
\begin{equation}
\pi(\lambda/\lambda_0) \simeq \expb{- \lambda/\lambda_0}, 
\elabel{exponential_free_path}
\end{equation}
even in the solid phase (see \fig{freepath}). This exponential
ansatz allows us to estimate the mean-square displacement for the
local Metropolis algorithm, supposing, for simplicity, that the
proposed moves have fixed step size $\ell$  in random directions. The
acceptance probability $p_{\acc}(\ell) = \expb{-\ell/\lambda_0}$
yields $\mean{\Delx^2( \ell)} = \ell^2 \expb{-\ell/\lambda_0}$, which
is maximized when $\ell=2\lambda_0$,
\begin{equation}
\max_\ell \mean{\Delx^2(\ell)} = 
\mean{\Delx^2(2 \lambda_0)} =  4\lambda_0^2 / \text{e}^2.
\elabel{maximal_local_metropolis}
\end{equation}

To estimate the mean-square displacement for the event-chain algorithms,
we suppose that the lengths of subsequent displacements in the chain
are independent. In this case, the expected number of particles in the
chain equals $\ell/\lambda_0 +1$.  We index
the displacement during one event-chain move through a time-like variable~$s$ with
$0\le s\le \ell$. The mean-square displacement of an event-chain move
(the expected sum of the squares of the individual displacements) can
be expressed through the probability $\pi(s,s')$ that two times $s$
and $s'$ belong to the same particle movement:
\begin{equation*}
\mean{\Delx^2(\ell)}= \int_0^\ell \int_0^\ell \dd{s}  \dd{s'} \pi(s,s').
\elabel{}
\end{equation*}
With the ansatz of \eq{exponential_free_path}, we have $\pi(s,s')=
\expb{-|s-s'|/\lambda_0}$. This yields the mean-square displacement
per particle, which can be viewed as a short-time (local) diffusion coefficient:
\begin{equation}
\Dloc^\infty(\ell)= \frac{\mean{\Delx^2(\ell)}}{\mean{M(\ell)}}= 2 \lambda_0^2
\underbrace{\frac{\exp(-\ell/\lambda_0)+\ell/\lambda_0-1}{\ell/\lambda_0+1}}_
      {\text{$\to 1$ for $\ell/\lambda_0 \to \infty$}}.
\elabel{mean_square_event}
\end{equation}
This tends to $2\lambda_0^2$ for large $\ell/\lambda_0$, that is,
to a value  $\text{e}^2 /2 \sim 4$ times larger than what we found
in \eq{maximal_local_metropolis}, for the local Metropolis algorithm.
This factor corresponds to the efficiency gain we might expect for a
generic event-chain algorithm with large $\ell / \lambda_0$, even though
we will obtain considerably larger factors for the \SEC\ algorithm.

In a finite system, the expressions in \eq{mean_square_event} must be
corrected for the center-of-mass displacement. For the \SEC\ algorithm,
the corrected mean-square displacement per particle, $\Dloc(\ell)$,
drops to zero for $\ell/\lambda_0 \to \infty$  because in that limit,
for a finite box, all disks participate in the chain, and the system ends
up moving as a solid block.  The \REC\  algorithm, in contrast, saturates
to a constant mean-square displacement per particle for large chains.

\begin{figure}[htbp]
\psfrag{-0.5}[cr][cr]{$-0.5$}
\psfrag{0.5}[cr][cr]{$0.5$}
\psfrag{0}[cc][cc]{$0$}
\psfrag{1}[cr][cr]{$1$}
\psfrag{10p8}[cc][cc]{$10^8$}
\psfrag{RePsi}[cc][cc]{$\operatorname{Re}\Psi$}
\psfrag{ImPsi}[cc][cc]{$\operatorname{Im}\Psi$}
\psfrag{Deltat}[cc][cc]{$\Delta_t$}
\psfrag{C(Deltat)}[cr][cr]{$C(\Delta_t)\!\!\!$}
\psfrag{exp(-Deltat/tau)}[cr][cr]{$\exp(\!-\!\Delta_t/\tau)\!\!\!$}
\psfrag{t}[cr][cr]{$_t$}
\psfrag{Y}[cr][cr]{$\Psi$}
\psfrag{D}[cr][cr]{$\Delta$}
   \centerline{\includegraphics{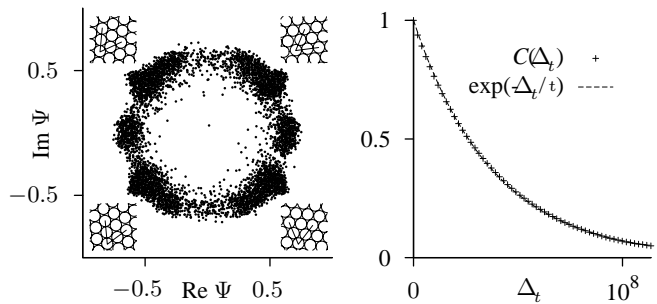}}
   \caption
   {\emph{Left}: Order-parameter distribution for $256$ disks in a
   periodic square box for $\eta=0.71$.
   \emph{Right}: Correlation function  $C(\Delt)$ for this system.
   The correlation time is obtained from an exponential fit, as shown. }
   \flabel{param_256}
\end{figure}

To further analyze the event-chain algorithms, we determined the
auto-correlation time of the orientational order parameter $\Psi$
\cite{strandburg} for hard-disk systems at densities near the melting
transition.  The orientational order parameter $\Psi$ averages the
complex-valued local bond order $\psi_j$ for each disk $j$, where
\begin{equation}
\Psi = 1/N \sum_j \psi_j
\elabel{order_parameter}
\end{equation}
and
\begin{equation}
\psi_j=\frac{1}{n_j} \sum_{\mean{k,j}} \expb{i 6 \varphi_{j,k}}.
\elabel{order_parameter_local}
\end{equation}
In \eq{order_parameter_local}, the sum is over the $n_j$ neighbors of
$j$, and $\varphi_{j,k}$ is the angle of the shortest vector equivalent
to $\xvec_k - \xvec_j$ \cite{strandburg}.  Probable values of
$\Psi$ form an irregular ring around the origin (see the scatter plot
in \fig{param_256}; the  $\Psi \to \Psi + \pi$ symmetry in a square box
imposes $\mean{\Psi(t)}=0$).

We suppose that the correlation time of this system is proportional to
the time the order parameter takes to wander around the ring, that is,
the auto-correlation time of~$\Psi$. This global measure of the overall
speed of a Monte Carlo simulation is more appropriate than, for example,
single-particle diffusion constants. However $\Psi$ is very long to
decorrelate at the interesting densities (see  \fig{scaling_tau}), and
we have to limit ourselves to small systems. The auto-correlation function
$C(\Delt)$ of the orientational order parameter is defined by the ensemble average
\begin{equation*}
C(\Delt) \propto \mean{\Psi(t) \Psi^*(t+ \Delt)}, 
\end{equation*}
normalized so that $C(0)=1$. In our square box,  this function decays
to zero exponentially for large $\Delt$ (see \fig{param_256}), and the
decay constant~$\tau$ and the speed $1/\tau$ are obtained by a fit,
for each value of the parameters $(N,\eta,\ell)$, from one single very
long simulation (with $t \gg \tau$). The local Metropolis algorithm,
for its optimal step size, is as fast as the event-chain algorithms with
$\ell/\lambda_0 \simeq 1$.  (Our implementation moves $3\times 10^{10}$
particles per hour on a 2.8GHz single-processor workstation for $N=4096$.)

\begin{figure}[htbp]
   \centerline{\includegraphics{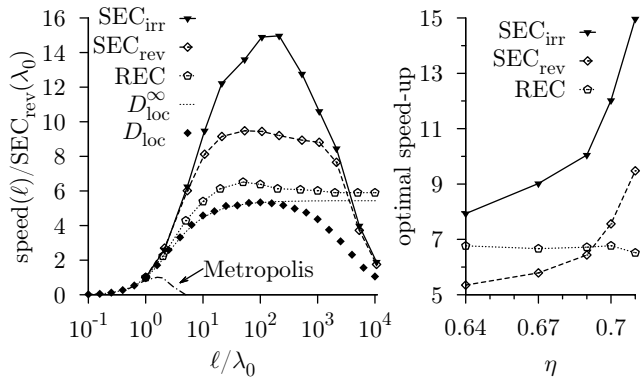}}
   \caption {\emph{Left}: Efficiency of \SEC\ and \REC\
  algorithms for $1024$ disks at $\eta=0.71$ (all speeds normalized
  by the speed of the reversible \SEC\ algorithm at $\ell/ \lambda_0 =
  1 $). The speed of the local Metropolis algorithm and the mean-square
  displacement per particle from \eq{mean_square_event} are also shown.
  \emph{Right}: Density dependence of the optimal speed-up factor.}
   \label{f:efficiency_1024}
\end{figure}

For small total displacements $\ell/\lambda_0 \ll 1$, the speed of
all the algorithms (reversible and irreversible \SEC, \REC, and local
Metropolis algorithm) is proportional to $\Dloc^\infty(\ell)$, as given
by \eq{mean_square_event}, that is, they all follow the single-particle
behavior (see \fig{efficiency_1024}).  For larger $\ell/\lambda_0$, the
event-chain algorithms realize a considerable speed-up compared to the
local Metropolis algorithm (also in \fig{efficiency_1024}).  Moreover,
both versions of the \SEC\ algorithm set up coherent motion across the
system and are clearly better than the \REC\ algorithm, whose particle
chains meander through the system (as shown in \fig{event_chain_moves}),
so that the disks move incoherently.  For large $\ell/\lambda_0$,
the irreversible \SEC\ algorithm is faster than the reversible version:
it is of advantage to break detailed balance.  \figg{efficiency_1024} also
illustrates that the \SEC\ algorithm becomes more efficient (as compared
to the local Metropolis algorithm) as one approaches the transition from
the liquid phase (at density $\eta \sim 0.708$). The optimal speed-up
increases with the system size, as shown in \tab{table1}.  This suggests
that the speed-up of the \SEC\ algorithm may well increase with the
correlation length of the system, and may, in the transition region,
have a more favorable scaling than the local Metropolis algorithm.

\begin{table}[htbp]
\centerline{\begin{tabular}{r|c|c}
    & \multicolumn{2}{c}{Optimal speed-up} \\
$N$ & Reversible & Irreversible \\
\hline
$64$ &  $\sim 6$ & $\sim 8$ \\
$256$ &  $\sim 8$ & $\sim 11$ \\
$1024$ &  $\sim 9$ & $\sim 15$ \\
$4096$ & $\sim 10$  & $\sim 20$
\end{tabular}}
\caption{Optimal speed-up reached by the \SEC\ algorithm (with respect
to the reversible \SEC\ algorithm for $\ell/\lambda_0=1$) at density
$\eta=0.71$ as a function of particle number.}
\tlabel{table1}
\end{table}

Let us finally discuss the relationship between the Monte Carlo method
and the molecular-dynamics algorithm. All these approaches describe
the same equilibrium state. Unlike the Monte Carlo method, the molecular
dynamics follows the physical time-evolution of the system. The
first implementations of the molecular dynamics algorithm
\cite{alder} were very time consuming, with a complexity of $O(N)$ per event
(collision), slower than the Metropolis algorithm ($O(1)$ per
move). The complexity of modern implementations has improved to
$O(\log N)$ \cite{Isobe} per event and even $O(1)$ \cite{paul}. A closer
look is thus needed to choose between the two methods. 

We used a simple version of the molecular dynamics
to compute the decorrelation time of $\Psi$ in the same way as in
\fig{param_256}. In number of events, molecular dynamics is found to
be about three times faster than the irreversible version of \SEC\ for
$\eta \sim 0.7$ and $N=64-1024$. It is very 
interesting to notice that molecular dynamics shows, unlike \REC ,
the same density dependence of its speed as \SEC\ around the transition
region. We then determined the CPU time per collision of one of the
most rapid current implementations of the hard-disk molecular-dynamics
algorithm \cite{Isobe}. For the $32\times 32$ system
at $\eta \sim 0.7$, this implementation reaches
about $1.7 \times 10^9$ collisions per hour on a $2.6$GHz workstation
\cite{Isobe_private}.  Our $xy$ implementation of the \SEC\ algorithm
reaches $3\times 10^{10}$ collisions per hour on similar
hardware. Our implementation is thus about $5$ times faster in CPU-time to reach
thermodynamic equilibrium than the best molecular-dynamics
implementation. We should also note that \SEC\ is much easier to implement.
A synopsis of these relative and absolute timing issues is presented 
in \fig{scaling_tau}. For clarity, we give times in terms of
\quot{equivalent Metropolis moves}, this means that one event of the
molecular dynamics algorithm corresponds to $\sim 3$ \SEC\ events and to $\sim
60$ Metropolis moves. The horizontal lines indicate what can be
achieved in approximately one hour with our implementation of the
Metropolis algorithm, irreversible \SEC, and the implementation of the
molecular dynamics algorithm of \cite{Isobe}.

\begin{figure}[htbp]
   \centerline{\includegraphics{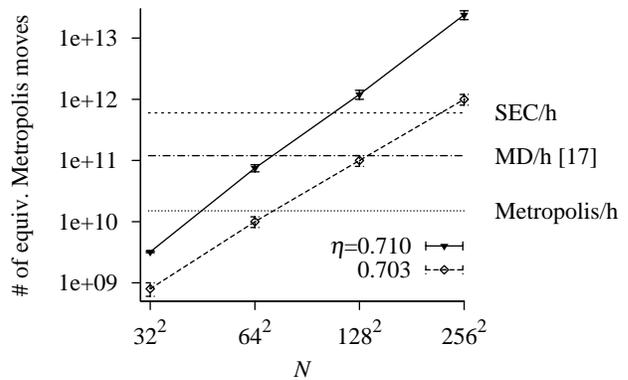}}
   \caption {System-size dependence of the
    correlation time of the orientational order parameter for two
    densities. What can be achieved in approximately 
    one hour using the different algorithms discussed in this paper is
    indicated by horizontal lines.}
   \flabel{scaling_tau}
\end{figure}

In conclusion, we have in this paper proposed a class of algorithms
for hard spheres and related systems, which clearly outperform the local
Metropolis algorithm. We discussed three aspects of our algorithms,
which all contributed to improve their speed. First, we showed that
event-chain algorithms have a larger effective step size than the local
Metropolis algorithm, because spheres move until they strike one of
their neighbors. We computed mean-square displacements per particle
(local diffusion constants) to quantify this point.  Nevertheless,
local diffusion constants are not clearly related to the speed of the
algorithm: they merely describe the short-time rattling of a particle
in its cage (only for small $\ell/\lambda_0$ is the local diffusion
constant directly proportional to the algorithm's speed). Second,
we performed numerical simulations of two variants of the method, and
carefully analyzed the auto-correlation function of the orientational
order parameter. One of them, the \SEC\ algorithm, induces coherent
motion of a long chain of spheres, and it allows the different parts
of the system to communicate with each other. We witnessed
considerable performance gains of this algorithm in the critical
region, in the same way than the molecular dynamics. This suggests the exciting
possibility that the speed-up of the event-chain algorithm grows with
the correlation length of the system, and may have a more favorable
scaling than the local Metropolis algorithm in the critical
region. This speed-up, which is shared by both the molecular-dynamics
algorithm and the \SEC\ algorithm, is not understood and should be further investigated.
Third, we noticed that the absence of rejections permitted to conceive
an irreversible version of the \SEC\ algorithm which improves the performances.

Our implementation of the \SEC\ algorithm approaches equilibrium (for large systems
at $\eta \simeq 0.70$) about 40 times faster than our local
Metropolis algorithm, not only because of the speed-up evidenced
in \fig{efficiency_1024} but also because the $xy$ version of the
algorithm computes no scalar products and uses very few random
numbers. It also equilibrates about five times faster than the best
molecular-dynamics implementation and preserves certainly a large
potential for improvement.

Nevertheless, CPU times needed for convergence remain extremely large,
and even with our algorithm, full convergence of systems with $10^6$
particles at high densities comes barely into reach. The irreversible \SEC\ algorithm not
only appears to be the fastest currently known simulation method for
dense hard-disk and hard-sphere systems, but it also provides a telling
example of the benefits of breaking detailed balance in Monte Carlo
algorithms going beyond the \quot{lifting} Markov chains \cite{MR1789978}.

\end{document}